\documentclass[twocolumn]{NobArticle}
\usepackage[ruled]{algorithm2e} 

\runninghead{D. Li et al.}

\title{RaRa Clipper: A  Clipper for Gaussian Splatting Based on Ray Tracer and  Rasterizer}

\author{
    Da Li,
    Donggang Jia,
    Yousef Rajeh,
    Dominik Engel,
    and Ivan Viola
    \\
    \texttt{\{da.li, donggang.jia, yousef.rajeh, dominik.engel, ivan.viola\}@kaust.edu.sa} \
}

\date{
    King Abdullah University of Science and Technology \\ 
}

\renewcommand{\maketitlehookd}{%
\begin{abstract}
    With the advancement of Gaussian Splatting techniques, a growing number of datasets based on this representation have been developed. However, performing accurate and efficient clipping for Gaussian Splatting remains a challenging and unresolved problem, primarily due to the volumetric nature of Gaussian primitives, which makes hard clipping incapable of precisely localizing their pixel-level contributions. In this paper, we propose a hybrid rendering framework that combines rasterization and ray tracing to achieve efficient and high-fidelity clipping of Gaussian Splatting data. At the core of our method is the RaRa strategy, which first leverages rasterization to quickly identify Gaussians intersected by the clipping plane, followed by ray tracing to compute attenuation weights based on their partial occlusion. These weights are then used to accurately estimate each Gaussian's contribution to the final image, enabling smooth and continuous clipping effects. We validate our approach on diverse datasets, including general Gaussians, hair strand Gaussians, and multi-layer Gaussians, and conduct user studies to evaluate both perceptual quality and quantitative performance. Experimental results demonstrate that our method delivers visually superior results while maintaining real-time rendering performance and preserving high fidelity in the unclipped regions.
    
    \medskip

    \small{\textbf{Keywords:} Clipping, Gaussian Splatting, Rasterization, Ray tracing.}
\end{abstract}
}

\begin{document}

\small
\maketitle

\section{Introduction}
With the rapid advancement of 3D vision and neural rendering, Gaussian Splatting (GS)~\cite{kerbl20233d} has emerged as a powerful representation for real-time rendering and differentiable reconstruction. Unlike traditional discrete representations such as point clouds, meshes, or voxel grids, Gaussian primitives offer continuous spatial support, anisotropy, and sub-pixel anti-aliasing, enabling both high fidelity and efficient rendering. As a result, this representation has been widely adopted in state-of-the-art research and open-source systems.

\begin{figure}[h]
    \centering
    \includegraphics[width=\linewidth]{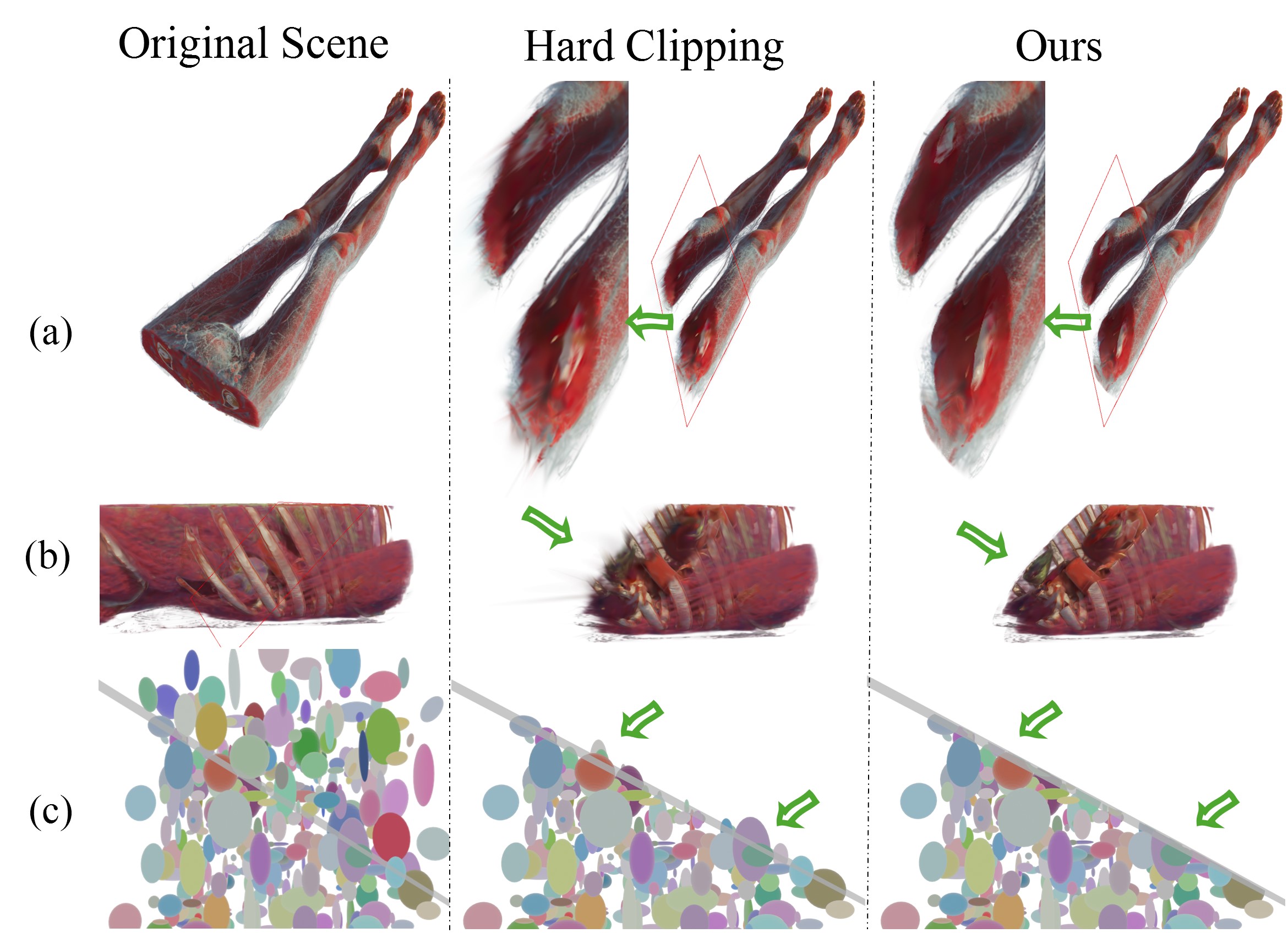}
    \caption{ Comparison of different clipping strategies across representative Gaussian Splatting scenes. (a) Lower body dataset(a) and full body dataset(b): our method preserves anatomical structure without boundary artifacts, while hard clipping introduces visible distortions. (c) Schematic visualization using ellipsoids highlights the effect of clipping at the primitive level—RaRa selectively attenuates partially visible Gaussians, avoiding abrupt removal.}
    \label{fig: fake_teaser}
\end{figure}

In many practical applications, such as interactive editing~\cite{tang2025ivr}, medical image clipping~\cite{kleinbeck2025multi}, robotic navigation~\cite{chen2024survey}, and hair modification~\cite{GSHair2024human,zhougroomcap}, it is often necessary to perform clipping operations to isolate regions of interest, remove occlusion, or accelerate rendering. While traditional graphics pipelines provide mature solutions such as z-buffering, frustum culling, and AABB-based rejection, these techniques~\cite{shirley2009fundamentals} are not directly applicable to Gaussian Splatting.

\begin{figure}
    \centering
    \includegraphics[width=\linewidth]{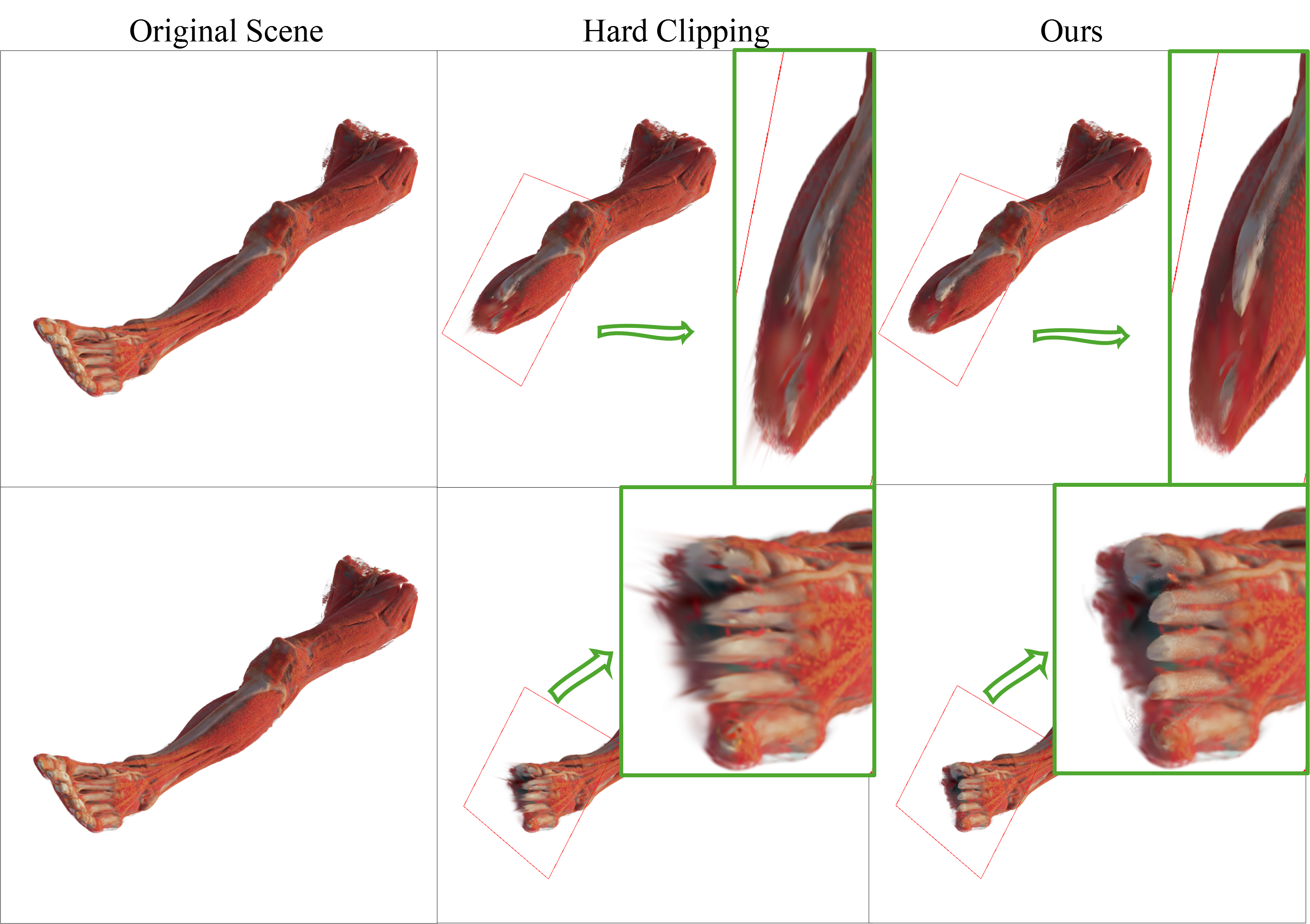}
    \caption{Comparison of clipping results on the full leg dataset using different clipping strategies and clipping planes. The dataset is clipped by two different planes. Our method preserves anatomical detail and structural continuity near the clipping boundary, while hard clipping introduces visible distortions and blurring artifacts.}
    \label{fig:full_leg_render}
\end{figure}

\begin{figure}
    \centering
    \includegraphics[width=\linewidth]{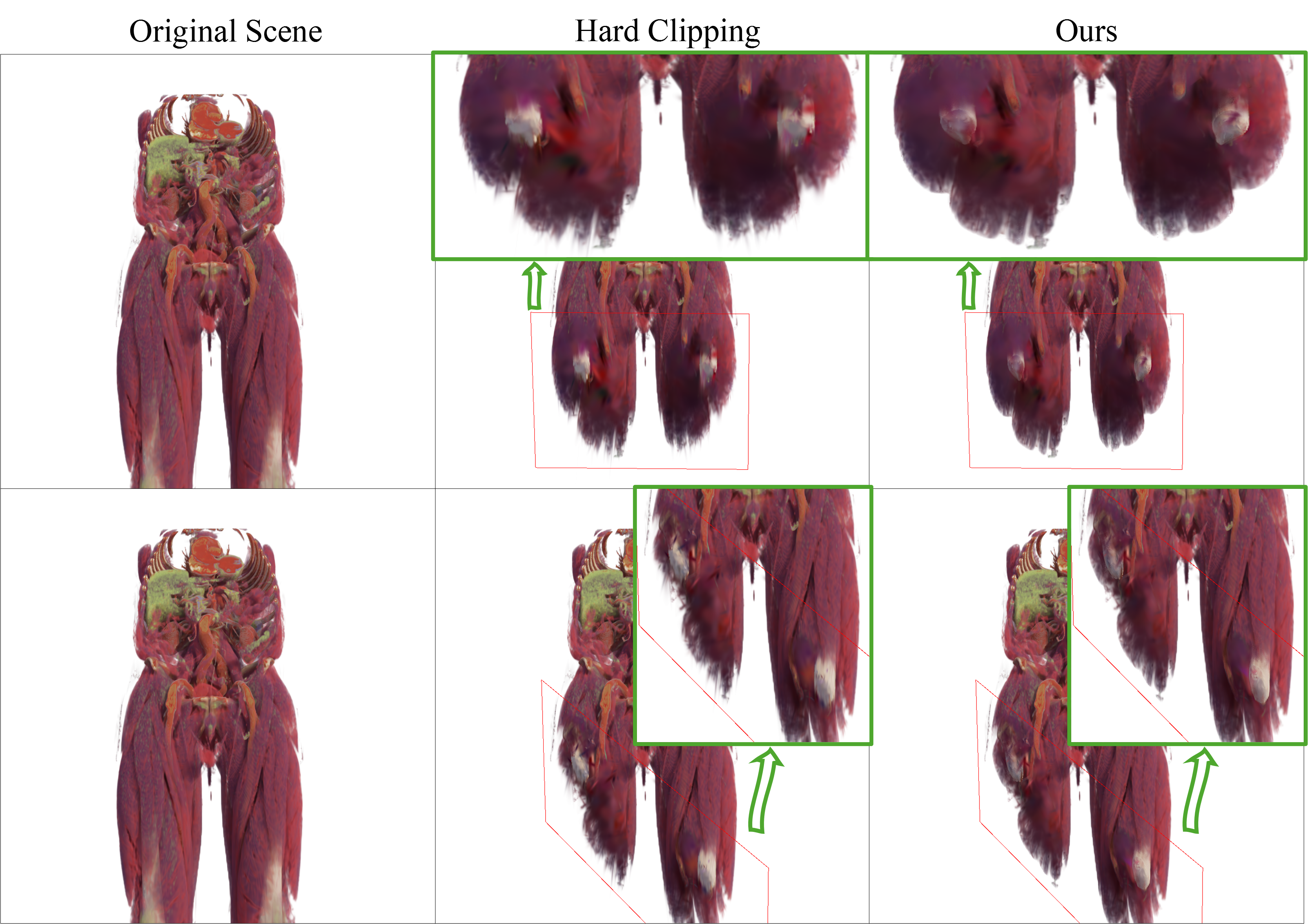}
    \caption{Clipping comparison on the full body dataset using different clipping strategies and clipping planes. Our method preserves internal structures and soft tissue continuity across boundaries, whereas hard clipping introduces aliasing and visible distortions.}
    \label{fig:full_body_render}
\end{figure}

\begin{figure}
    \centering
    \includegraphics[width=\linewidth]{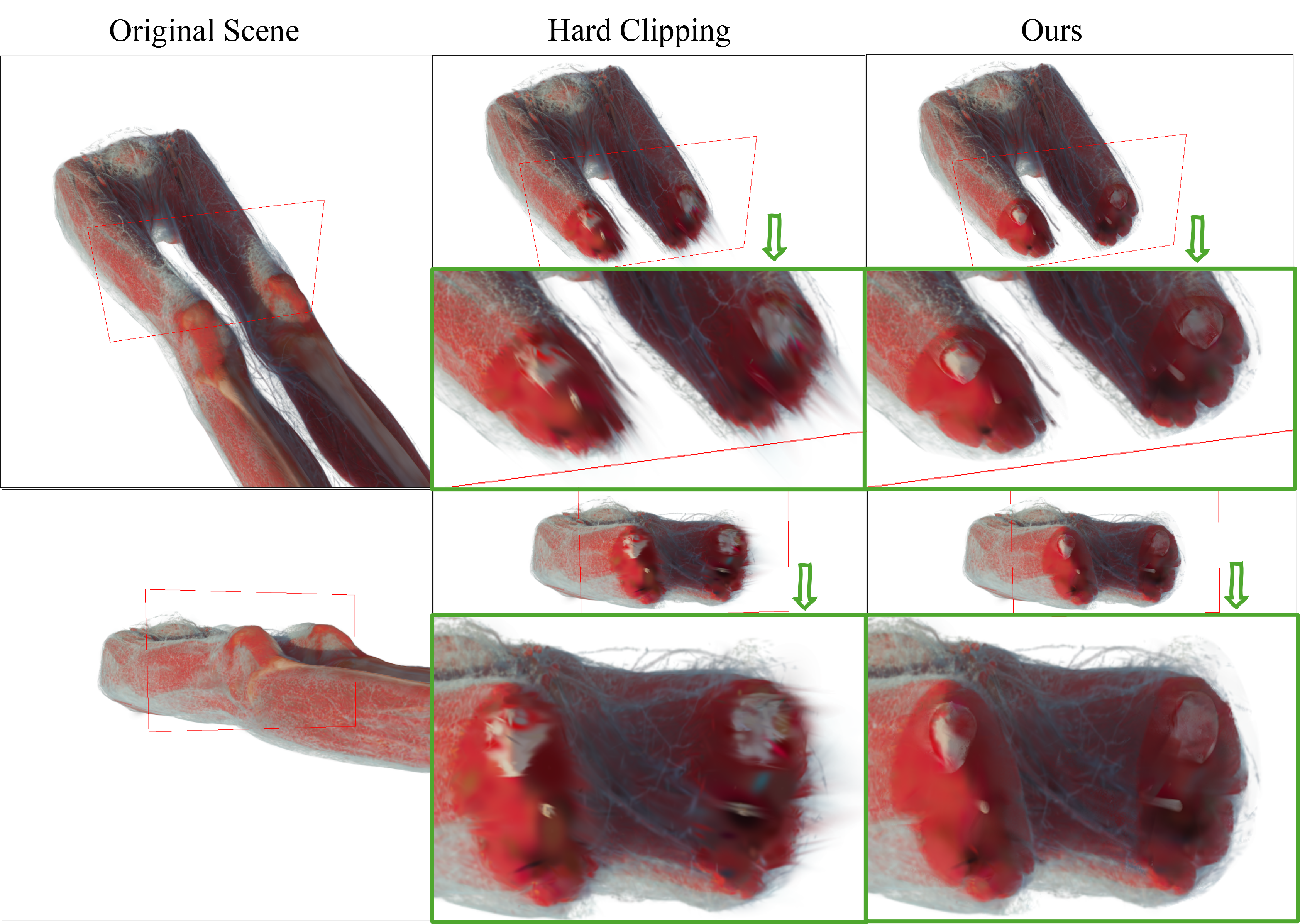}
    \caption{Comparison of clipping results on the lower body dataset using different clipping strategies and views. Each row presents a different camera viewpoint, but under the same clipping plane configuration. Our method effectively preserves structural integrity and surface smoothness near the clipping interface, while hard clipping produces visible artifacts such as aliasing and detail loss.}
    \label{fig:lower_body_render}
\end{figure}

\begin{figure}
    \centering
    \includegraphics[width=\linewidth]{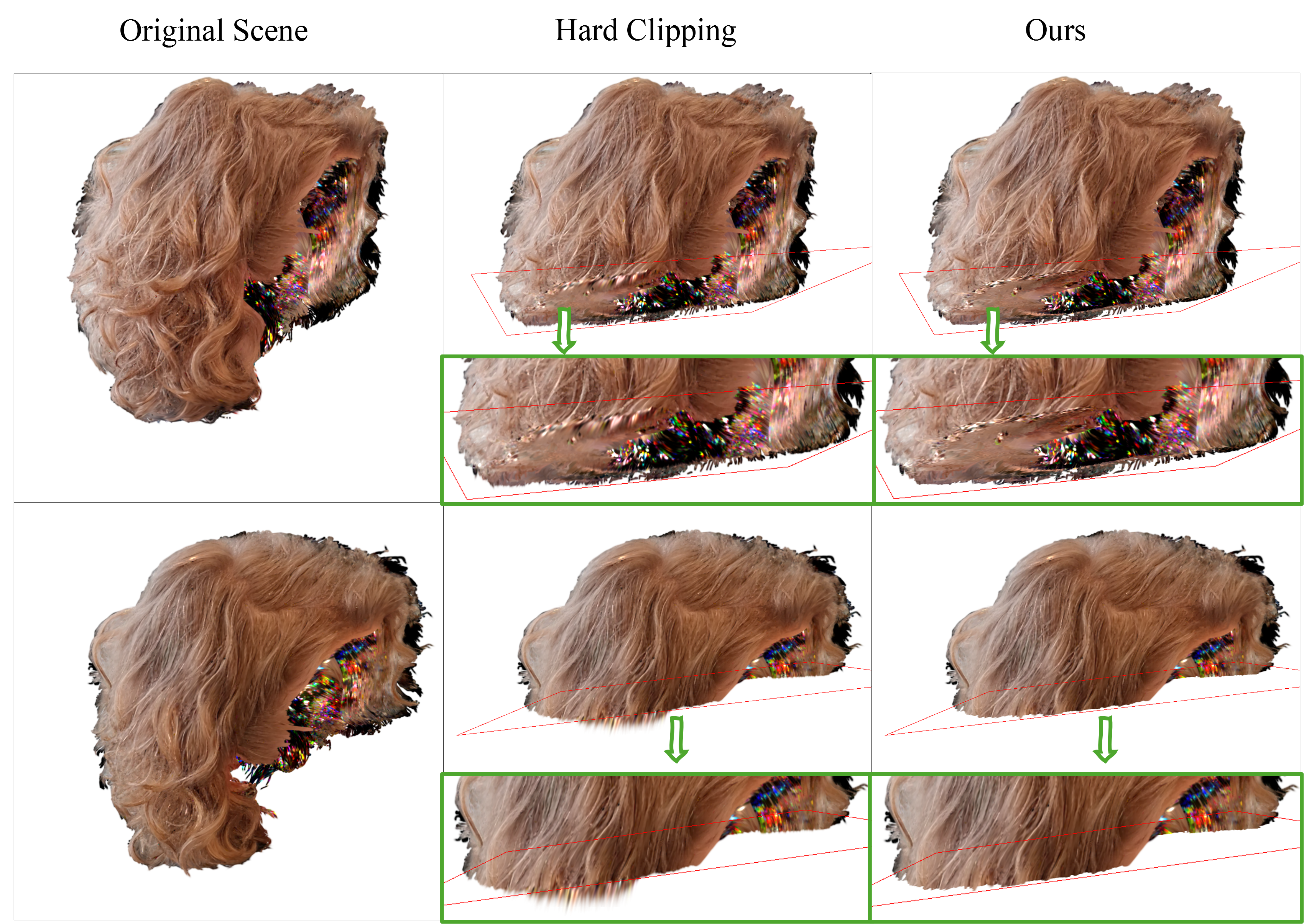}
    \caption{Clipping comparison on the Hair dataset using different clipping strategies. Our method eliminates artifacts along the clipping boundary, while the hard clipping approach introduces noticeable spiky distortions at the ends of hair strands. Each row corresponds to a different clipping plane configuration. 
}
    \label{fig:hair_render}
\end{figure}

\begin{figure}
    \centering
    \includegraphics[width=\linewidth]{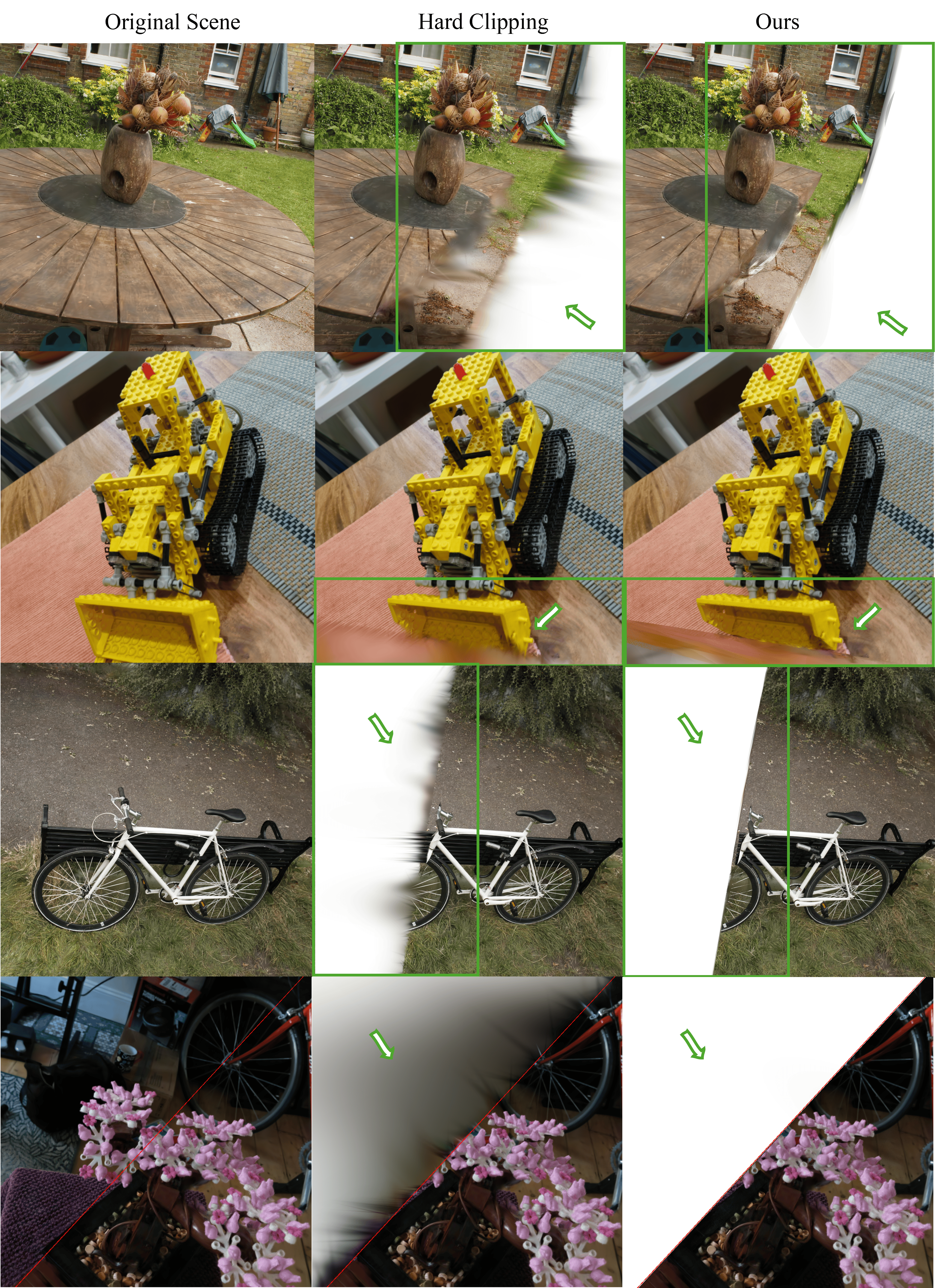}
    \caption{Clipping results on general appearance modeling datasets, showcasing the impact of different strategies near the clipping boundary. Hard clipping introduces spiky artifacts, particularly visible in edge structures. In contrast, our method produces clear contour cutting. The last row shows results on the 360bonsai dataset, where the red line indicates the visualized clipping plane. 
}
    \label{fig:general_scene_render}
\end{figure}

The primary challenge lies in the volumetric nature of Gaussian primitives: their contribution to image formation is computed by integrating their density along viewing directions, rather than through surface-based geometry. Existing clipping strategies~\cite{kerbl20233d,kleinbeck2025multi} employ a binary filtering approach based solely on the position of the Gaussian center. In this paper, we refer to this as the \textbf{Hard Clipping} strategy. However, this method fails to account for the true image-space contribution of each Gaussian, often leading to visual artifacts such as aliasing, popping, and the unintended removal of partially visible primitives, shown in \autoref{fig: fake_teaser}, ~\autoref{fig:full_leg_render}, and Figure ~\ref{fig:full_body_render} -~\ref{fig:general_scene_render}.


To address these issues, we propose a novel \textbf{Ra}y-\textbf{Ra}sterizer (\textbf{RaRa}) clipping framework that combines the parallel visibility analysis capabilities of rasterization with the precise intersection reasoning of ray tracing. Our method consists of two key stages:  (1) we classify the visibility of each Gaussian by computing its signed distance to the clipping plane, labeling it as fully visible, fully invisible, or near the clipping boundary (cutoff); and (2) we apply ray tracing only to Gaussians in the cutoff region to compute ellipsoid-ray intersections and derive a decaying opacity weight based on the length of the visible ray segment.


This selective ray tracing avoids unnecessary computation, reduces errors caused by hard clipping, and maintains structural continuity across clipped regions. Moreover, the proposed strategy is general and can be integrated into any system that supports Gaussian Splatting.

We evaluate our approach on three representative datasets---general Gaussians scenes, strand Gaussians scenes~\cite{GSHair2024human,zhougroomcap,luo2024gaussianhair}, and multi-layer Gaussians scenes~\cite{tang2025ivr,kleinbeck2025multi}---and conduct both quantitative and perceptual evaluations via user studies. Results demonstrate that our method consistently preserves visual fidelity, shown in \autoref{fig: fake_teaser} and real-time performance.

Our main contributions can be summarized as follows:
\begin{itemize}
    \item We propose the RaRa hybrid clipping framework—the first to integrate rasterization and ray tracing for efficient and precise Gaussian Splatting clipping, which can be seamlessly integrated as an add-on to a general Gaussian Splatting renderer.
    \item We design a physically-consistent decaying opacity function based on visible ray segment length, enabling accurate modeling of partial occlusion.
    \item We conduct comprehensive experiments and user studies on multiple Gaussian datasets, demonstrating superior visual quality and robustness compared to hard clipping.
\end{itemize}

\section{Related Research}
\subsection{Gaussian Splatting and Its Applications}
Gaussian Splatting has recently gained significant attention as a continuous 3D representation technique for real-time rendering and differentiable scene reconstruction. The seminal work by Kerbl et al.~\cite{kerbl20233d} introduced 3D Gaussian Splatting, which uses anisotropic Gaussian primitives as rendering units and employs efficient image-space rasterization without requiring explicit meshes similar to previous work~\cite{westover1991splatting,zwicker2001ewa}. Compared to point-based~\cite{kivi2022real} or neural implicit methods~\cite{mildenhall2021nerf,barron2021mip}, GS provides higher reconstruction fidelity while maintaining interactive performance.

Subsequent research has explored various applications and extensions of GS, including sparse-view reconstruction~\cite{liu2025review}, dynamic scene reconstruction~\cite{duan20244d,lin2024gaussian,yang2024deformable}, avatar reconstruction~\cite{qian20243dgs,li2024animatable,pang2024ash,hu2024gauhuman}, and multi-scale modeling~\cite{li2024mipmap,seo2024flod,ren2024octree}. Highly anisotropic Gaussians have been applied to fine-grained strand-level hair modeling~\cite{GSHair2024human,zhougroomcap,luo2024gaussianhair}, and anatomical data has been converted to layered Gaussian fields~\cite{kleinbeck2025multi} for efficient rendering on consumer VR devices~\cite{niedermayr2024application}. Works like iVR~\cite{tang2025ivr} further introduced editable Gaussian attributes and multi-layer fusion, enhancing interaction and structural representation. These studies demonstrate the versatility of GS beyond appearance modeling, indicating a growing need for standardized operations such as clipping, which remains underexplored and motivates this paper.

\subsection{Limitations of Traditional Clipping in Gaussian Representations}
Clipping and culling are standard operations in traditional graphics pipelines, primarily using techniques such as view frustum culling and Z-buffer depth testing~\cite{shirley2009fundamentals,tran1986fast}. These methods perform well on rigid, surface-based geometry like meshes or point clouds. However, Gaussian primitives are volumetric in nature and provide soft spatial support defined by covariance matrices, making them incompatible with conventional binary visibility tests.

Some systems attempt hard clipping by measuring the distance from the Gaussian center to the clipping plane, but such heuristics often produce artifacts—flickering, popping, or structural loss—especially near clipping boundaries. For example, the original 3DGS~\cite{kerbl20233d} includes basic frustum-based visibility culling based on Gaussian centers, and Multi-layer GS~\cite{kleinbeck2025multi} implements bounding box clipping but still relies on center-based hard culling logic. Other domain-specific works include ellipsoidal volume clipping on the digital earth platform using voxel-aligned slicing~\cite{wang2023blh}, fast proxy-based clipping of subterranean networks~\cite{konev2018fast}, and image-space clipping in textured volume rendering \cite{weiskopf2003interactive}. These methods, however, operate on voxel grids, regular proxy geometry, or raster-aligned volumes, and cannot be directly applied to the soft-support, anisotropic primitives used in Gaussian Splatting.

Thus, implementing an efficient and accurate clipping framework for Gaussian fields remains an open problem. The core difficulty lies in accounting not only for binary visibility but also for the integral contribution of partially visible Gaussians to each pixel.

\subsection{Ray Tracing for Gaussian Rendering}
To enhance photorealism and physical modeling in GS-based rendering, recent works have introduced ray tracing into Gaussian rendering pipelines. For instance, 3DGRT~\cite{moenne20243d} utilizes bounding volumes for efficient ray–Gaussian intersection and supports reflection/refraction modeling. 3DGUT~\cite{wu20243dgut} retains rasterization but integrates ray tracing for performance optimization. REdiSplats~\cite{byrski2025redisplats} approximates Gaussians as mesh-based flat shapes, enabling standard mesh-editing operations. Relightable 3D Gaussians~\cite{gao2024relightable} combine ray tracing with BRDF decomposition for realistic relighting. Stochastic Ray Tracing~\cite{sun2025stochastic} adopts efficient sampling strategies to accelerate transparency modeling in Gaussian scenes.

Although these techniques significantly improve global rendering quality and light transport modeling, they rarely address localized clipping or fine-grained visibility estimation. Many require custom ray tracing pipelines or mesh priors, which increase training costs or reduce rendering performance~\cite{mai2024ever}. In contrast, our RaRa strategy focuses on real-time, localized clipping by selectively applying ray tracing only to cutoff Gaussians. Rasterization is first used to identify potentially clipped Gaussians, followed by ray–ellipsoid intersection and visible volume ratio computation. Our approach is plugin-style and can be directly integrated into native GS pipelines without altering training or data structures, enabling precise and interactive clipping at negligible runtime overhead.

\section{Method}
In the rendering process of Gaussian Splatting, we need to project each 3D Gaussian distribution onto the 2D image plane to compute its contribution to pixel color. However, due to the infinite support of the Gaussian function, we typically limit the effective range of influence to within three standard deviations. Although this constraint improves rendering efficiency, it still cannot explicitly represent the complete point set on the ellipsoidal surface. As a result, standard depth-buffer-based rasterization is incapable of accurately determining whether a Gaussian intersects with the clipping region. Our method addresses this shortcoming using ray-tracing for the clipped Gaussians. 

In this section, we first introduce Gaussian Splatting formally (Sec.~\ref{sec:gs-preliminaries}), before presenting our approach to facilitate clipping during rendering. This approach begins by identifying the Gaussians that can be trivially rendered or discarded, because they are completely on one side of the clipping plane (Sec.~\ref{sec:gaussian-threeway}). Next, we focus on the Gaussians that are intersected by the clipping plane. For these Gaussians, we compute their exact intersection with the clipping plane (Sec.~\ref{sec:intersection}) and adapt their appearance based on this intersection using a decaying function (Sec.~\ref{sec:decaying-function}). The full method is summarized in \autoref{alg:RaRaClip}.

\subsection{Gaussian Splatting}\label{sec:gs-preliminaries}
3D Gaussian Splatting~\cite{kerbl20233d} is a framework for 3D reconstruction using point clouds with Gaussian kernels. Each point is represented by a Gaussian centered at $\boldsymbol{\mu} \in \mathbb{R}^3$ with a covariance matrix \( \mathbf{V} \) which can be represented by the scale and rotation matrix as \( \mathbf{V}= \mathbf{R} \mathbf{S}^2 \mathbf{R}^T \). The Gaussian kernel is defined as:
\begin{equation}
G(\mathbf{x},\boldsymbol{\mu},\mathbf{V}) = \frac{1}{(2\pi)^{3/2} |\mathbf{V}|^{1/2}} \exp\left(-\frac{1}{2} (\mathbf{x} - \boldsymbol{\mu})^\top \mathbf{V}^{-1} (\mathbf{x} - \boldsymbol{\mu})\right).
\end{equation}

Each point has attributes such as transparency \( \alpha_k \), opacity density $\delta_k$, and color $c_k$. $\alpha_k$ is calculated by $\delta_k$, shown in Sec.~\ref{sec:decaying-function}.  The rendering process involves projecting the point cloud into camera space and performing splatting operations to generate the final image. To blend color attributes accurately, points are sorted by depth, and pairwise compositing is applied to account for transparency. The compositing formula is:
\begin{equation}
C = \sum_{k \in N} c_k \alpha_k\prod_{j=1}^{k-1} (1 - \alpha_j),
\end{equation}


\subsection{Identifying non-trivial Gaussians}\label{sec:gaussian-threeway}
\begin{figure}[h]
    \centering
    \includegraphics[width=0.75\linewidth]{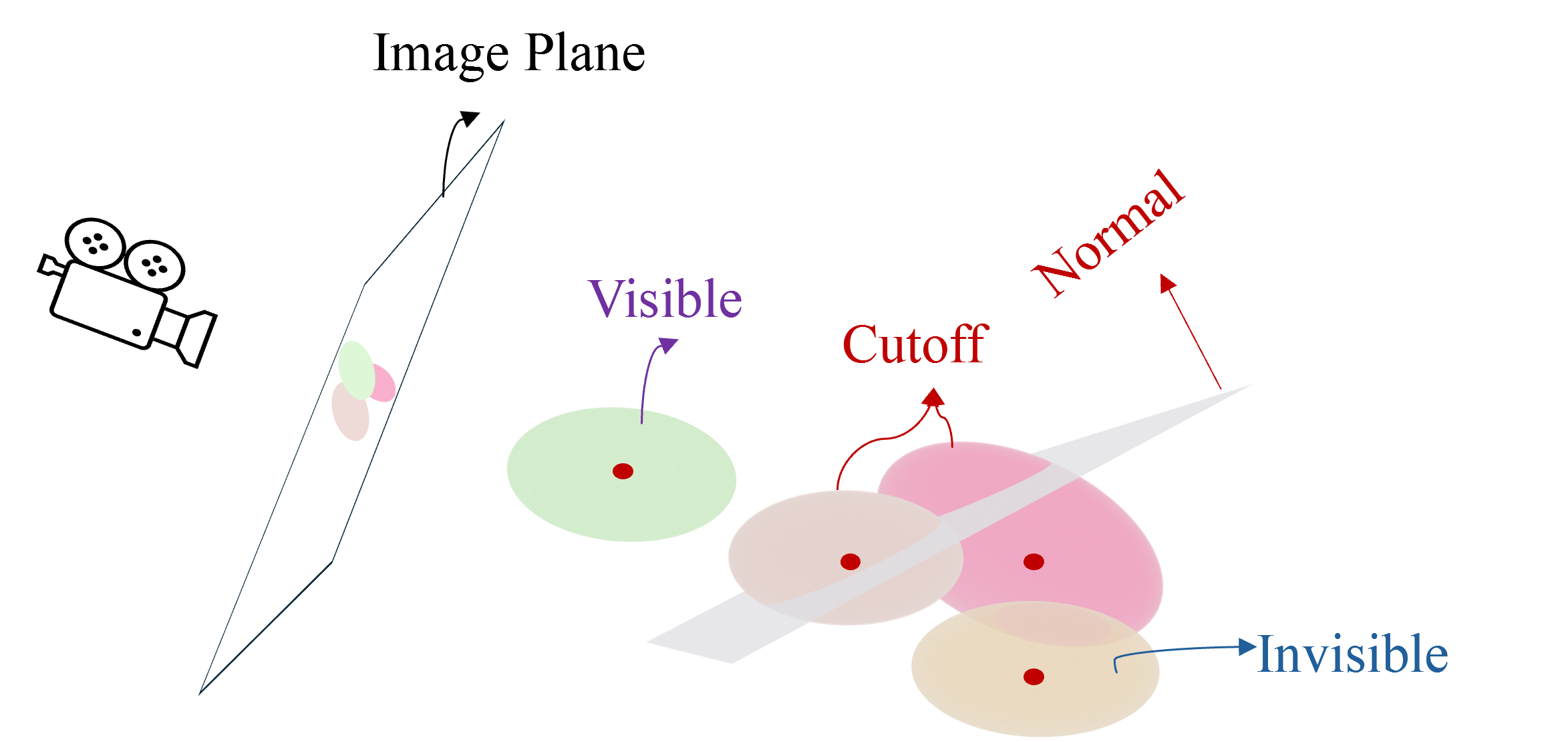}
    \caption{Multiple Gaussians classified in different categories in the scene. The visible region is defined as the half-space in the direction of the clipping plane’s normal.}
    \label{fig: multiGS}
\end{figure}
In order to clip these Gaussian representations, we first aim to categorize the set of Gaussians, to decide if they will be discarded, rasterized, or ray-traced, based on where their centers lie in comparison to the clipping plane. Our goal is to restrict expensive ray-tracing to the Gaussians that cannot be trivially handled through rasterization or discarding.
To distinguish the Gaussians, we begin by defining the clipping region using an implicit plane:
\begin{equation}
\pi(\mathbf{x}) = \mathbf{n}\mathbf{x} + d = 0
\end{equation}
where $\mathbf{n} \in \mathbb{R}^3$ is the normal of the clipping plane and $d \in \mathbb{R}$ is the signed distance from the origin to the plane. 
In a scene containing multiple Gaussians, as illustrated in \autoref{fig: multiGS}, where we define the area the normal pointing to is the visible region for the viewer, each Gaussian $G_k$ with center $\boldsymbol{\mu}_k$ can be classified into one of three categories based on its signed distance to the clipping plane:

\begin{enumerate}[leftmargin=5ex] 
\item \textbf{Invisible:} If the center is far behind the clipping plane, i.e., $d_{e.p.} = \mathbf{n}\boldsymbol{\mu}_k + d < -r_k$, where $r_k = 3  \max(\sigma_{x_k}, \sigma_{y_k}, \sigma_{z_k})$, the Gaussian is discarded. $\sigma$ is the standard deviation. \( d_{e.p.} \) denotes the signed distance from the Gaussian center (ellipsoid) to the clipping plane.
\item \textbf{Visible:} If the center lies clearly in front of the plane, i.e., $d_{e.p.} > r_k$, the Gaussian is fully visible and rasterized normally.
\item \textbf{Cutoff:} If the center is within the proximity of the clipping boundary, i.e., $|d_{e.p.}| \leq r_k$, the Gaussian is partially visible. We compute precise intersections and modify the appearance.
\end{enumerate}
Note that some Gaussians could be falsely classified in the cutoff region, as we ignore anisotropy by using the maximum $\sigma$ for this classification.
To reduce computational cost and keep rendering fidelity, we limit exact intersection tests to Gaussians in the cutoff category. These are the only cases where visibility is ambiguous and visual adaptation of the Gaussians is required.

\subsection{Ray-Gaussian and Clipping Plane Intersection}\label{sec:intersection}

To enable precise clipping of Gaussian ellipsoids, we now derive the exact intersections of a viewing ray with the clipped Gaussians, as well as the clipping plane.
To compute the intersection with the Gaussian, we model the transformed Gaussian using a linear transformation from the unit sphere. Specifically:

\begin{itemize}
\item An ellipsoid is represented as $E(\mathbf{S}, \mathbf{R}, \boldsymbol{\mu})$, where $\mathbf{S}$ is the scale matrix, $\mathbf{R}$ is the rotation matrix, and $\boldsymbol{\mu}$ is the center.
\item The transformation matrix $\mathbf{M} = \mathbf{R} \mathbf{S}$ maps the unit sphere to the ellipsoid, and the transformed point is given by $\mathbf{x} = \mathbf{M} \mathbf{u} + \boldsymbol{\mu}$ with $\mathbf{u} \in \mathbb{S}^2$, where \( \mathbb{S}^2 \) represents the unit sphere embedded in \( \mathbb{R}^3 \).
\end{itemize}

Consider a ray $\mathbf{r}(t) = \mathbf{e} + t  \mathbf{d}$, where $\mathbf{e}$ is the ray origin, $\mathbf{d}$ is the normalized direction vector, and $t \geq 0$. Transform this ray into the local space of the unit sphere:
\begin{equation}
\tilde{\mathbf{r}}(t) = \mathbf{M}^{-1}(\mathbf{e} - \boldsymbol{\mu}) + t\mathbf{M}^{-1} \mathbf{d}
\end{equation}
We then solve for the intersection of this ray with the unit sphere using:
\begin{equation}
|\tilde{\mathbf{r}}(t)|^2 = 1
\end{equation}
which leads to the quadratic equation:
\begin{equation}
At^2 + Bt + C = 0
\end{equation}
With coefficients:
\[
\left\{
\begin{aligned}
A &= \|\mathbf{M}^{-1} \mathbf{d}\|^2 \\
B &= 2 \langle \mathbf{M}^{-1}(\mathbf{e} - \boldsymbol{\mu}), \mathbf{M}^{-1} \mathbf{d} \rangle \\
C &= \|\mathbf{M}^{-1}(\mathbf{e} - \boldsymbol{\mu})\|^2 - 1
\end{aligned}
\right.
\]
Solving this quadratic gives the two intersection parameters (for simplicity, here we only consider the case of two intersection points):
\begin{equation}
t\_e_{1,2} = \frac{-B \pm \sqrt{B^2 - 4AC}}{2A}
\end{equation}
Substituting $t\_e_1$ and $t\_e_2$ into $\mathbf{r}(t)$ gives the two world-space intersection points with the transformed ellipsoid.

Lastly, we compute the intersection of the ray with the clipping plane to determine if and where the Gaussian is actually clipped. To achieve this, we solve for the intersection of the ray with the clipping plane $\pi(x) = 0$. 
Substituting $\mathbf{r}(t)$ into the plane equation yields:
\begin{align}
t\_p &= \frac{-(\mathbf{n}  \mathbf{e} + d)}{\mathbf{n} \mathbf{d}}
\end{align}
If the resulting $t\_p$ is between $t\_e_1$ and $t\_e_2$, the ray intersects the ellipsoid and is also clipped by the plane. Otherwise, the Gaussian is either fully visible or fully occluded and treated accordingly. In the following section, we will use these intersection points to determine how clipped Gaussians need to be adjusted visually.

\subsection{Decaying Function Design}\label{sec:decaying-function}

In the standard rendering formulation of Gaussian Splatting, each 3D Gaussian splat contributes to pixel color by integrating its projected density along the ray direction. For any position in the space  $\mathbf{x} \in \mathbb{R}^3$, and its corresponding pixel coordinate $\mathbf{\hat{x}} \in \mathbb{R}^2$ in the image space, we can compute the opacity contribution for this pixel of any Gaussian points $\boldsymbol{\mu}_k \in \mathbb{R}^3$ by~\cite{zwicker2001ewa}:

\begin{equation}
\label{eq:gaussian_projection}
\alpha_k(\mathbf{\hat{x}}) = \delta_k\int_{\mathbb{R}} \frac{1}{|\mathbf{J}^{-1}| |\mathbf{W}^{-1}|} G\left( \mathbf{x},\boldsymbol{\mu}_k,{\mathbf{J} \mathbf{W} \mathbf{V}_k \mathbf{W}^{T} \mathbf{J}^{T}} \right) dx_z
\end{equation}
where $\delta_k$ indicates the opacity density, $\mathbf{J}$ denotes the Jacobian matrix of the projective transformation, and $\mathbf{W}$ is the transformed matrix from world to camera space. $V_k$ represents the 3D Gaussian covariance matrix in world space, while $G(\cdot)$ is the 3D Gaussian kernel function. $\hat{\boldsymbol{\mu}}_k$ denotes the projection of the Gaussian center in the image space, and $\mathbf{x_z}$ is the projection direction.

We can directly get the analytical result of Equation~\ref{eq:gaussian_projection} as:
\begin{equation}
\label{eq:simplified_gaussian_projection}
\alpha_k(\mathbf{\hat{x}}) = \frac{1}{|\mathbf{J}^{-1}||\mathbf{W}^{-1}|} \delta_k \hat{G} \left( \hat{\mathbf{x}}, \hat{\boldsymbol{\mu}}_k,{\hat{\mathbf{V}}_k} \right),
\end{equation}
where $\hat{G}$ is the 2D Guassian kernel, $\hat{\mathbf{V}}_k$ is obtained by skipping the last row and column of $\mathbf{J} \mathbf{W} \mathbf{V}_k \mathbf{W}^{T} \mathbf{J}^{T}$.


\begin{figure}[h]
    \centering
    \includegraphics[width=0.75\linewidth]{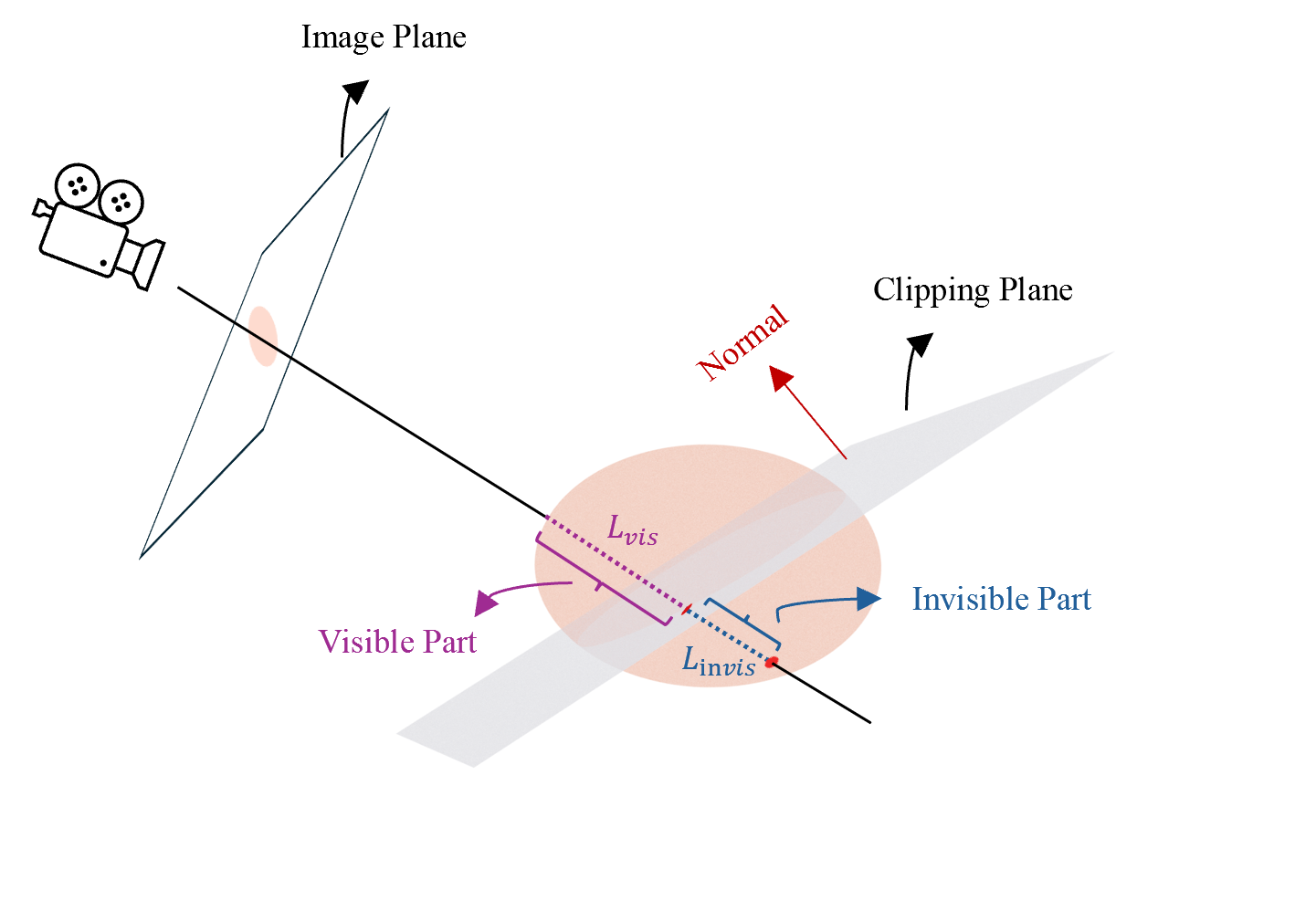}
    \caption{Illustration about calculation for decay weight for cutoff Gaussians.}
    \label{fig:Gaussian Clipping}
\end{figure}

When a Gaussian is partially removed by a clipping plane, like in~\autoref{fig:Gaussian Clipping}, we cannot use the analytic result shown in \autoref{eq:simplified_gaussian_projection} and need to compensate for the missing portion. To approximate the residual visible contribution, we propose a 1D length-ratio-based decay function $w(\mathbf{\hat{x}})$. The key idea is to estimate the proportion of the Gaussian volume contributing to opacity by measuring the length of the visible segment along the ray within the Gaussian. As shown in \autoref{fig:Gaussian Clipping}, we define $L_\text{vis}$ as the segment between ray–ellipsoid intersections that lies within the visible region, and $L_\text{invis}$ as the portion falling in the invisible region. The full extent of the intersection path through the ellipsoid is then given by $L_\text{vis} \cup L_\text{invis}$. We define the decay weight function as:
\begin{equation}
w(\mathbf{\hat{x}}) = 
\begin{cases}
\frac{L_\text{vis}}{L_\text{vis} \cup L_\text{invis}}, & \text{if two intersections exist} \\
1, & \text{otherwise (keep original rasterization)}
\end{cases}
\end{equation}
Based on the decay weight $w(\mathbf{\hat{x}})$, we can then update the opacity of the Gaussian to adjust its contribution to the pixel:
\begin{equation}
\alpha_k^{\text{new}} = \alpha_k(\mathbf{\hat{x}})w(\mathbf{\hat{x}})
\end{equation}
This adjusted opacity is then used in the final splatting step. This approach ensures that clipping operations do not introduce visible discontinuities or flickering along the edges.
The core logic of the full RaRa strategy is elucidated in \autoref{alg:RaRaClip}.

\begin{algorithm}[]
\caption{RaRaClip: Gaussian Visibility and Decaying Weight Computation}
\label{alg:RaRaClip}
\KwIn{Gaussian parameters $\mu_i, R_i, S_i$, clip plane ${(\mathbf{n}, d)}$,\newline camera pose $T$, image resolution $W \times H$, pixel coordinate $(p_x,p_y)$}
\KwOut{Opacity contribution $\alpha_i$}

\tcp{Step 1: Visibility Pre-check (Hard Clipping)}
$d{\text{vis}} \leftarrow \text{dot}(\mathbf{n}, \boldsymbol{\mu_i}) + d$; $r_i \leftarrow 3 \max(S_i)$; \\
\If{$d{\text{vis}} < -r_i$}{
\Return $\alpha_i \leftarrow 0$ \tcp*{Completely invisible}
}

\tcp{Step 2: RaRa Strategy - Ray Tracing for Cutoff}
\If{$|d_{\text{vis}}| \leq r_i$}{
$\mathbf{d} \leftarrow \text{ProjectRay}(p_x,p_y, T)$ \tcp*{pixel to ray direction}
$\mathbf{e} \leftarrow \text{CameraOrigin}(T)$

$(\tilde{\mathbf{d}}, \tilde{\mathbf{e}}) \leftarrow \text{TransformToUnitSphere}(\mathbf{d}, \mathbf{e}, M_{i}^{-1},\mu_i)$

$(t\_{e1}, t\_{e2}) \leftarrow \text{RayEllipsoidIntersect}(\tilde{\mathbf{d}}, \tilde{\mathbf{e}})$

$t_p \leftarrow \text{RayPlaneIntersect}(\mathbf{e}, \mathbf{d}, \mathbf{n}, d)$ \\
\If{$t\_e_1, t\_e_2 > 0$ and $t_p > 0$}{
$\mathbf{x}_1 = \mathbf{e} + t\_e_1 \mathbf{d}$, $\mathbf{x}_2 = \mathbf{e} + t\_e_2 \mathbf{d}$; \tcp*{Compute 3D Points}
\If{$\text{IsVisible}(\mathbf{x}_1, \mathbf{n}, d)$ \textbf{AND} $\text{IsVisible}(\mathbf{x}_2, \mathbf{n}, d))$ \textbf{AND} $t\_e_1 < t_p < t\_e_2$}{
$w\leftarrow 1.0$;
}
\ElseIf{$(\text{IsVisible}(\mathbf{x}_1, \mathbf{n}, d)$ \textbf{OR} $\text{IsVisible}(\mathbf{x}_2, \mathbf{n}, d)$}{
$w \leftarrow L_\text{vis} / (L_\text{vis} \cup L_\text{invis})$
}
}
}
\Else{$w \leftarrow 1.0$ \tcp*{Fully visible, keep opacity}}

\tcp{Step 3: Final Opacity Computation}
$\alpha_i^{\text{orig}}$ $\leftarrow$ \autoref{eq:simplified_gaussian_projection}

$\alpha_i \leftarrow \min(0.99, \alpha_i^{\text{orig}}w)$

\Return $\alpha_i$
\end{algorithm}

\section{Experiment}
In this section, we experimentally validate the effectiveness of our algorithm. We first introduce the evaluation datasets, then verify that our clipping approach retains the real-time performance of Gaussian Splatting. Ablation studies further reveal that naively applying ray tracing degrades quality, while our RaRa strategy effectively mitigates such artifacts. All experiments are conducted on an NVIDIA RTX 4090 GPU.
\subsection{Dataset}
We evaluate our method on three representative types of Gaussian datasets:

\textbf{General Gaussian scenes} are from SplatViz~\cite{barthel2024gaussian} framework using Mip-NeRF360, representing typical surface-level reconstructions.

\textbf{Strands Gaussian scene} is derived from the GaussianHair~\cite{GSHair2024human}, containing highly anisotropic primitives to showcase fine-scale effects such as hair clipping.

\textbf{Multi-layer Gaussian scenes} are based on anatomical reconstructions~\cite{kleinbeck2025multi} with layered Gaussian fields, enabling visualization of internal clipping within volumetric structures.




\subsection{Render Efficiency}
To ensure that our proposed clipping framework maintains real-time interactivity, we evaluate its runtime performance under dynamic clipping scenarios. To further verify the scalability of our method under dense scenes, we conduct evaluations on two large-scale datasets: 360garden ($\sim$4 million Gaussians) and Hair ($\sim$3 million Gaussians). Results in \autoref{tab:fps_results} show that our method achieves comparable frame rates to hard clipping and preserves real-time interactivity even in dense point-based representations. Original denotes the baseline rendering performance measured at a fixed camera view without any clipping. For both Hard Clipping and our RaRa Clipping method (Ours), we introduce a single clipping plane and gradually move it across the entire scene over the course of 30 frames, while keeping the camera view unchanged. The reported FPS values are averaged over these 30 frames to reflect interactive rendering performance under active clipping.

\begin{table}[h]
\centering
\renewcommand{\arraystretch}{1.2}
\begin{tabular}{lcccccc}
\toprule
& \multicolumn{3}{c}{\textbf{360garden} ($\sim$4M)} & \multicolumn{3}{c}{\textbf{Hair} ($\sim$3M)} \\
\cmidrule(lr){2-4} \cmidrule(lr){5-7}
\multirow{2}{*}{\textbf{FPS}} 
    & Original & Hard & Ours & Original & Hard & Ours \\
    \cmidrule(lr){2-2}
    \cmidrule(lr){3-3}
    \cmidrule(lr){4-4}
    \cmidrule(lr){5-5}
    \cmidrule(lr){6-6}
    \cmidrule(lr){7-7}
    & 77.82 & 81.49 & 81.47 & 128.02 & 146.88 & 145.46   \\
\bottomrule
\end{tabular}
\caption{Average rendering performance (FPS) under dynamic clipping.}
\label{tab:fps_results}
\end{table}

\subsection{Ablation Study}

To validate the necessity and correctness of the proposed RaRa strategy, we conduct an ablation study that compares rendering results with and without it. Our goal is to demonstrate that ray tracing should only be applied to Gaussians that lie near the clipping boundary, as naively tracing all Gaussians may introduce visual inconsistencies.

Since the exact ground truth of a scene after partial clipping is hard to define, we simulate a reference condition by placing the clipping plane at infinity. This ensures that no Gaussians are clipped and the entire scene remains fully visible. Under this assumption, the correct rendering should match that of the original unmodified Gaussian scene.

We compare the following two strategies:
\begin{itemize}
    \item \textbf{wo RaRa}: ray tracing and attenuation weights are performed on all visible Gaussians.
    \item \textbf{w RaRa}: ray tracing is only applied to Gaussians that are classified as \textit{cutoff}. 
\end{itemize}

\begin{figure}[h]
    \centering
    \includegraphics[width=\linewidth]{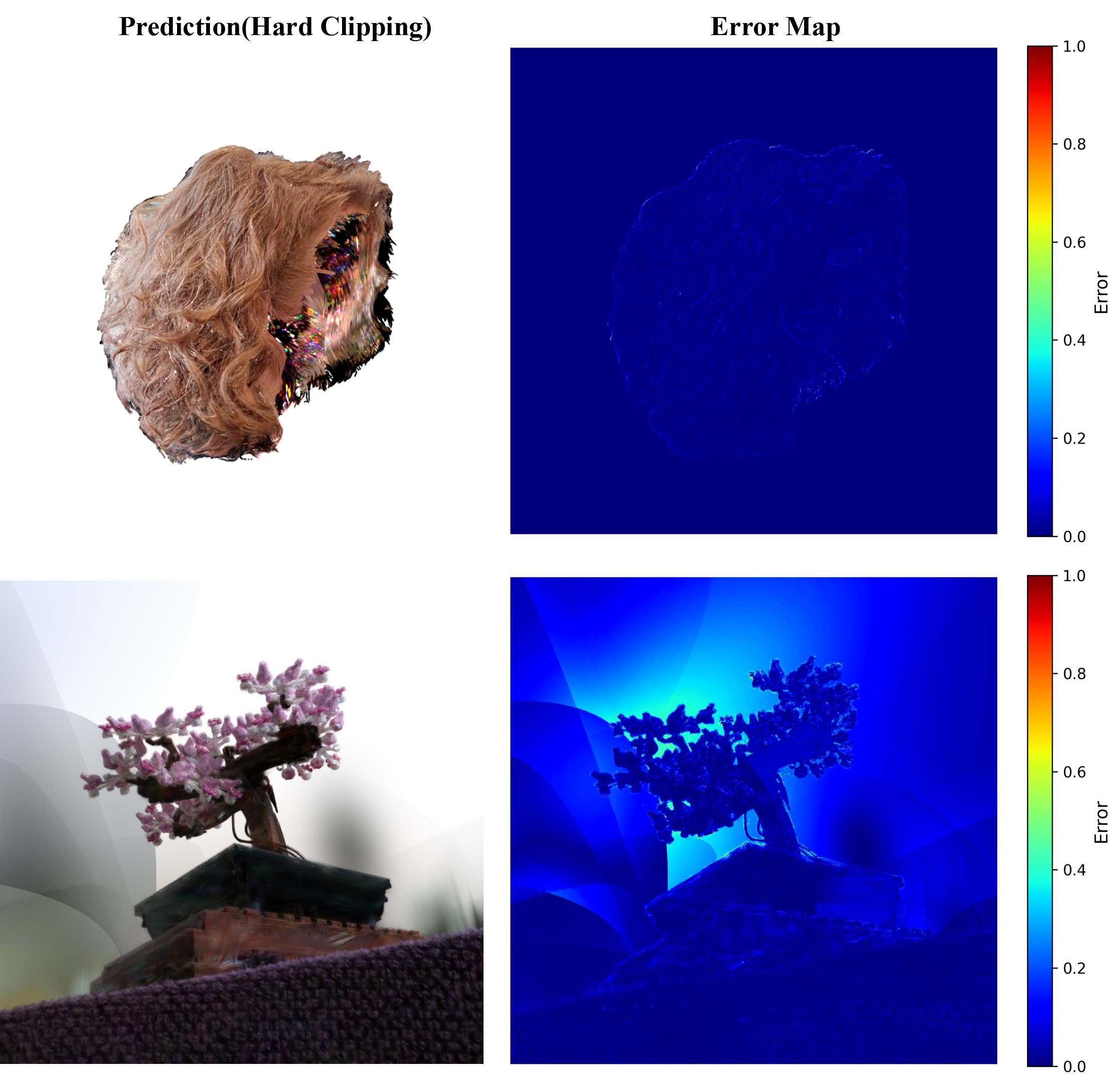} 
    \caption{Error visualization for the hard clipping strategy. Top row: hair dataset; Bottom row: 360bonsai dataset. We omit the RaRa strategy here, as it produces artifact-free rendering under our setup and yields zero error. }
    \label{fig:error_map}
\end{figure}

As shown in Table~\ref{tab:ablation_fps}, the RaRa strategy results in perfect alignment with the ground truth, with zero L1 error and SSIM of 1.0 on both datasets because no cutoff Gaussians exist. In contrast, without RaRa, the results exhibit numerical deviation and perceptual distortion, particularly along Gaussian boundaries as shown in \autoref{fig:error_map}.

\begin{table}[h]
\centering
\renewcommand{\arraystretch}{1.2}
\begin{tabular}{lcc@{\hskip 2em}cc}
\toprule
\textbf{} & \multicolumn{2}{c}{\textbf{360bonsai}} & \multicolumn{2}{c}{\textbf{Hair}} \\
\cmidrule(lr){2-3} \cmidrule(lr){4-5}
 & L1 & SSIM & L1 & SSIM \\
\midrule
wo RaRa & 8.5359 & 0.5759 & 0.3210 & 0.9968 \\
w RaRa  & \textbf{0.0} & \textbf{1.0} & \textbf{0.0} & \textbf{1.0} \\
\bottomrule
\end{tabular}
\caption{Ablation study comparing rendering results with and without the RaRa strategy under a simulated infinite clipping plane. Our method ensures perfect consistency with the ground truth.}
\label{tab:ablation_fps}
\end{table}

\section{User Study}

To systematically evaluate the user experience of our clipping strategies, we conducted a controlled user study comparing RaRa Clip—our proposed clipping method—with hard clipping. 
\subsection{Procedure}
The entire user study was carried out \emph{online} by our deployed website.  
Before starting the task, each participant read a concise task description as well as step-by-step instructions that covered the clipping techniques to be compared, the rating workflow, and all privacy guidelines.

\paragraph{Participants.}
We recruited a total of 22 participants (P1 to P22, 8 female, 14 male; mean age = 26.1 ± 3.9 years). In terms of academic background, 55\% held a Master’s degree, 27\% were PhD candidates or graduates, and 18\% held a Bachelor’s degree. Regarding computer-graphics experience, 55\% reported no prior CG experience, 23\% were beginners, 14\% intermediate users, and 9\% experts. This distribution offers a balanced mix of novices and specialists for evaluating our methods across varying expertise levels. They were rewarded with gift vouchers worth approximately 20 USD. 

\paragraph{Stimuli.} 
We collected \textbf{20 evaluation cases} drawn from \textbf{eight Gaussian splatting datasets}.  
Every case contained two strictly matched visualizations:
\begin{itemize}[noitemsep,leftmargin=12pt]
    \item A static rendering produced with the \emph{Hard Clip} method.
    \item A static rendering produced with our \emph{RaRa Clip} method.
    \item A short animation of the cutting plane sweeping through the datasets (shared by both methods).
\end{itemize}
Camera pose, cutting-plane orientation, and the plane’s start/end positions were kept \emph{identical} between the two methods, ensuring that the clipping algorithm was the \emph{only} varying factor.

\paragraph{Evaluation metrics.}
Each case was assessed on the following six subjective dimensions:
\begin{enumerate}[label=\arabic*.,noitemsep,leftmargin=18pt]
    \item \textbf{Flatness / Smoothness}: Degree to which the clipping surface is visually continuous and free of aliasing, stair-steps, or jagged curvature.
    \item \textbf{Temporal Stability}: The degree to which the rendered surface evolves smoothly and continuously as the clipping plane moves, without popping artifacts or abrupt changes in visibility that disrupt user perception.
    \item \textbf{Topological Integrity}:  Absence of unintended holes, breaks, or floating fragments that would disrupt anatomical or structural continuity.
    \item \textbf{Naturalness}: How seamlessly the clipped surface blends with surrounding textures, giving the impression of a ``natural dissection'' rather than an artificial cut.
    \item \textbf{Region of Interest (ROI) Exposure}: Extent to which the region of interest is revealed clearly, without over-occluding adjacent important structures.
    \item \textbf{Context Completeness}: The surrounding anatomy or environment is left visible to help users stay oriented.
\end{enumerate}

\paragraph{Fatigue control.}
Participants were randomly divided into two equally sized groups.  
Each group rated exactly \textbf{three metrics} per case, so that no participant completed more than~60 case–metric combinations, keeping the session within a comfortable time window.

\paragraph{Rating workflow.}
For every \emph{case–metric} combination, the following two-step procedure was performed:
\begin{enumerate}[label=\arabic*.,leftmargin=18pt]
    \item \textbf{Forced choice.}  
          The two results were displayed side-by-side in random order.  
          The participant selected the method that performed better on the current metric.
    \item \textbf{Strength rating.}  
          Immediately afterwards, the participant indicated \emph{how much} better the chosen method was, using a \textbf{7-point Likert scale} (1~=\;``barely better'', 7~=\;``extremely better'').
\end{enumerate}

\paragraph{Post-study questionnaire.}
After completing all rating tasks, participants answered three open-ended questions to gather qualitative insights:

\begin{enumerate}[label=\arabic*.,noitemsep,leftmargin=18pt]
    \item \emph{What key properties should an ideal clipping possess to support effective data exploration and interpretation?}
    \item \emph{What limitations or missing capabilities can be identified in the current two clipping methods?}
    \item \emph{What strengths or advantageous features do the current two clipping methods offer?}
\end{enumerate}

All interactions and timestamps were logged automatically for subsequent analysis.
\subsection{Result}
Across all six perceptual metrics, participants exhibited a clear and consistent preference for \textbf{RaRa Clip} over \textbf{Hard Clip}. Median ratings fell between~3 and~4 on the $[-7,7]$ scale, and Wilcoxon signed-rank~\cite {rosner2006wilcoxon} tests confirmed that every metric differed significantly from~0 in favor of \textbf{RaRa Clip} (all $p<.001$).

\begin{figure}
    \centering
    \includegraphics[width=\linewidth]{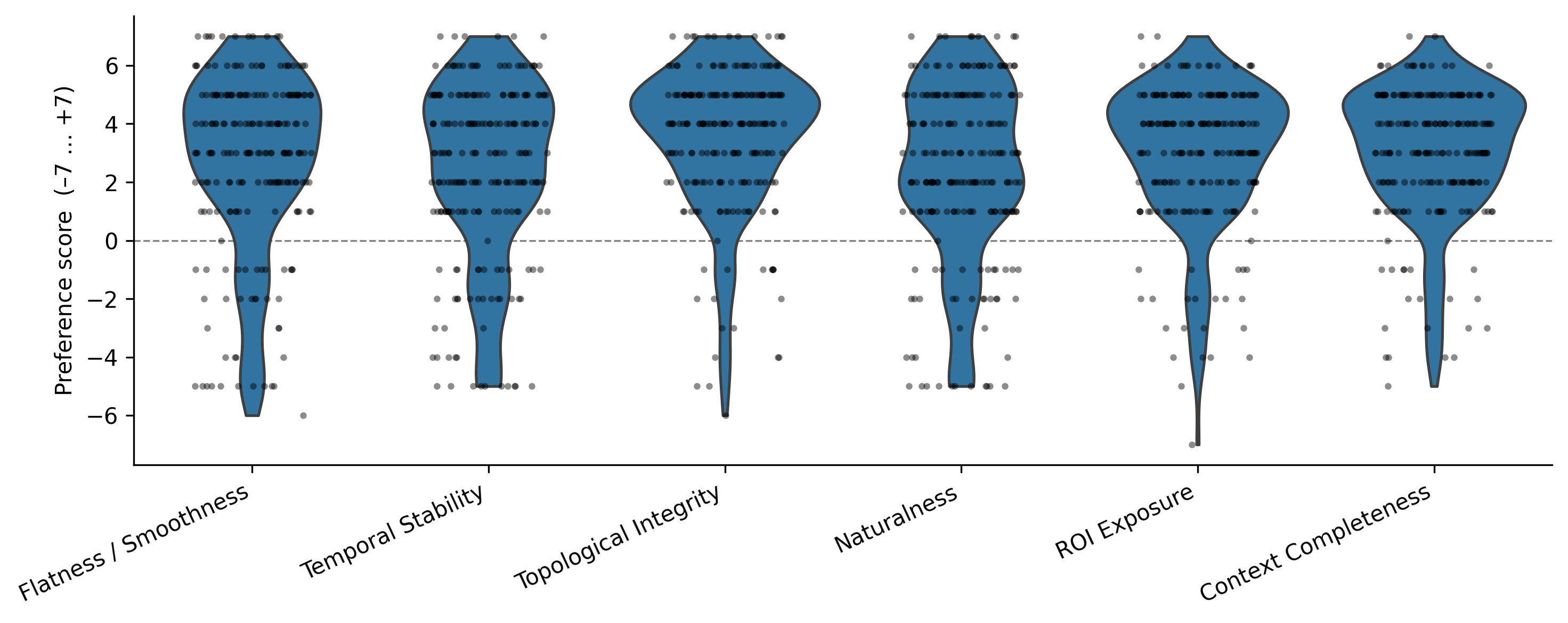}
    \caption{Violin plots users' ratings for the six metrics; positive scores favor RaRa Clip, revealing a clear overall bias toward it. Strip plots with a jitter in the figure show the original data amount.}
    \label{fig:violin}
\end{figure}

\begin{figure}
    \centering
    \includegraphics[width=0.6\linewidth]{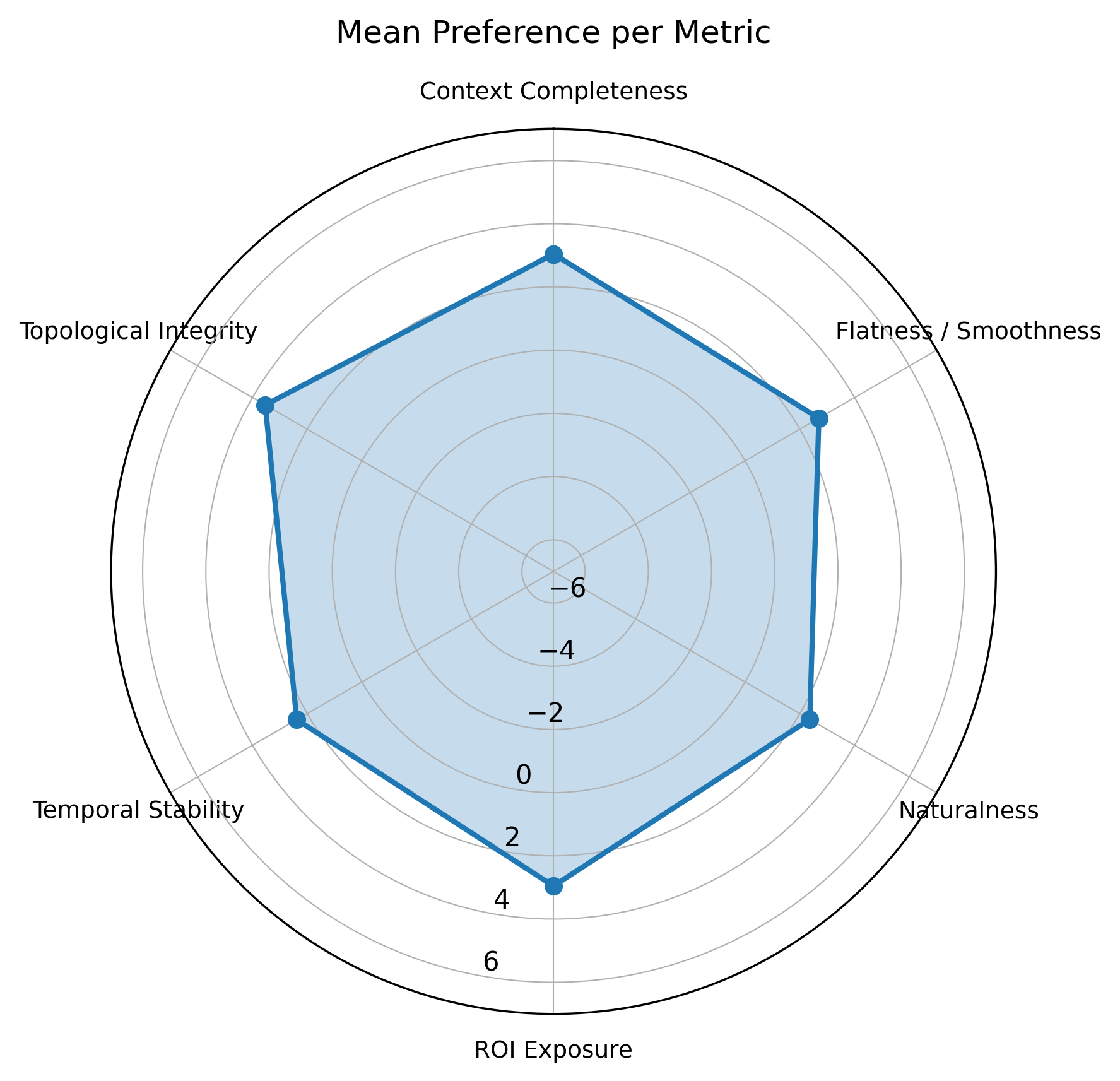}
    \caption{Radar chart of the mean preference scores for each of the six evaluation metrics; larger radii indicate stronger favor toward RaRa Clip, showing that participants consistently preferred it across all metrics.}
    \label{fig:radar}
\end{figure}

The largest effects emerged for \emph{Topological Integrity} ($M=3.53\pm2.47$, $d=1.43$, $91\,\%$ positive responses) and \emph{Context Completeness} ($M=3.04\pm2.28$, $d=1.33$, $91\,\%$ positive), indicating that users strongly valued preservation of structural relationships and surrounding context. Here, $M$ denotes the mean rating with standard deviation, and $d$ is the effect size~\cite{cohen2013statistical}, reflecting the magnitude of difference compared to the \emph{Hard Clip}. \emph{ROI Exposure} also showed a large effect ($M=2.95\pm2.44$, $d=1.21$, $90\,\%$ positive), suggesting that the proposed technique improves visibility of regions of interest without sacrificing context. The user rating results are demonstrated in the \autoref{fig:violin} and \autoref{fig:radar}.

Moderate, yet still substantial, benefits were observed for \emph{Flatness/Smoothness} ($M=2.70\pm3.06$, $d=0.88$, $82\,\%$ positive), \emph{Naturalness} ($M=2.36\pm3.05$, $d=0.78$, $81\,\%$ positive) and \emph{Temporal Stability} ($M=2.38\pm3.04$, $d=0.78$, $80\,\%$ positive). Across all metrics, the proportion of negative ratings never exceeded $20\,\%$, underscoring the robustness of user preference for \textbf{RaRa Clip}. Collectively, these results demonstrate statistically and practically significant gains in perceptual quality, with particularly strong advantages in maintaining topological fidelity, contextual completeness, and region-of-interest visibility.

\begin{figure}
    \centering
    \includegraphics[width=1\linewidth]{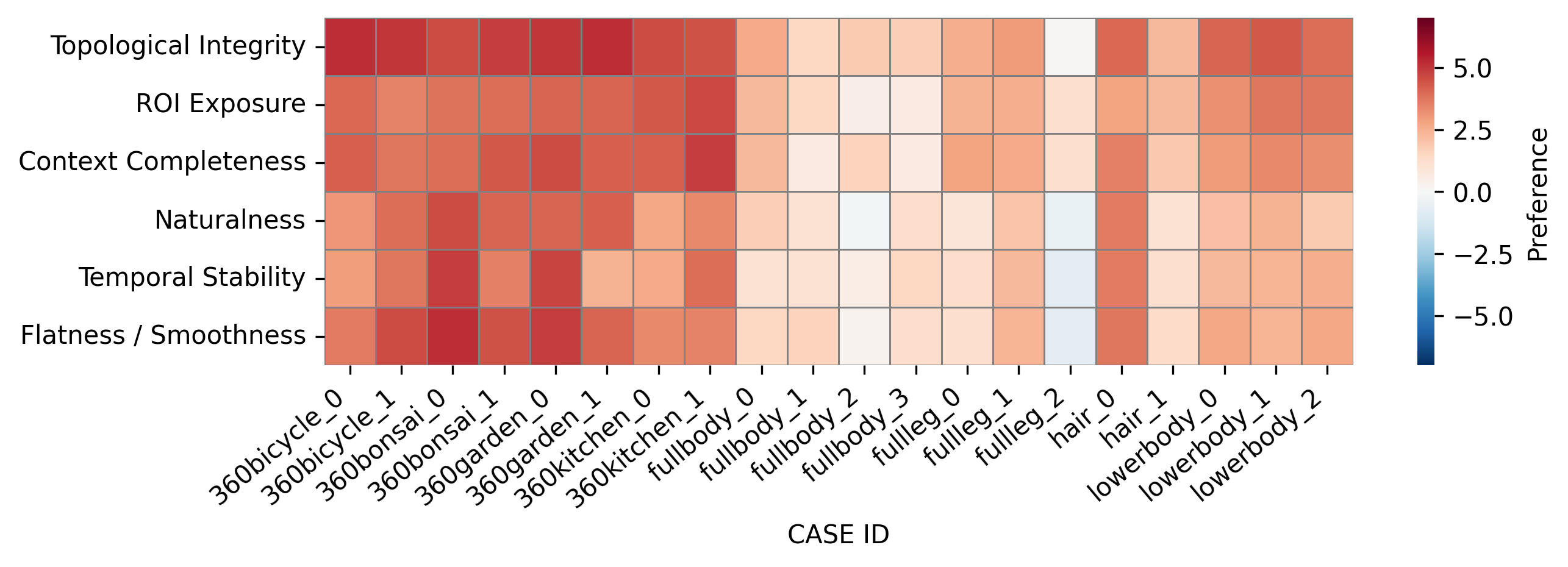}
    \caption{Heatmap of per-case preference scores for the six evaluation metrics; columns represent the 20 cases, rows the metrics, with blue cells indicating stronger favour toward Hard Clip and red cells toward RaRa Clip.}
    \label{fig:heatmap}
\end{figure}
We further conducted a case-by-case analysis across the six evaluation metrics, as shown in \autoref{fig:heatmap}. Red indicates a user preference for \textbf{RaRa Clip}, while blue indicates a preference for \textbf{Hard Clip}. The results show that, for nearly all cases and metrics, users consistently preferred RaRa Clip. An exception was observed in \textit{Case 2} of the \textit{full leg} dataset, where users rated Hard Clip higher in terms of \textit{naturalness}, \textit{temporal stability}, and \textit{flatness}. Additionally, the preference for RaRa Clip was more pronounced in real-world datasets such as \textit{bicycle}, \textit{bonsai}, and \textit{kitchen}, while the preference was slightly less evident in medical datasets.
\subsection{User Feedback}

To complement the qualitative results, we collected open-ended responses from participants regarding their preferences and expectations for effective clipping plane methods. Several key themes emerged:

\textbf{Smoothness and Visual Continuity.} Many users emphasized that an ideal clipping plane should be visually smooth and free of jagged edges or flickering artifacts. P3 specifically noted that one method maintained clear boundaries even during motion, while the other showed sudden changes or ghosting effects. P4 further pointed out that one method produced smoother transitions and fewer aliasing artifacts compared to another method.

\textbf{Precision and Low Distortion.} Users stressed the importance of preserving the geometric and perceptual fidelity of the clipped surface. This includes maintaining relative distances and orientations between data points to ensure the interpretability of the visualization. P13 explicitly likened the need for clarity to medical imaging, where preserving structure without overwhelming detail is critical.


\textbf{Dataset-dependent Preferences.} Participants (e.g., P6, P15, P16) observed that the effectiveness of each method varies by dataset. Viewpoint also played a role—differences between methods became more apparent when the cutting plane was not perpendicular to the camera.


\textbf{Advantages and Limitations.} Several users noted that both methods were accurate in general (P13, P21), but they also pointed out limitations such as insufficient background context preservation (P13) or visual inconsistencies near the clipping boundary (P11). 
\section{Discussion}

\textbf{ Underperforming scenarios}: While our RaRa clipping strategy consistently outperforms hard clipping across most datasets and metrics, we observed a notable outlier in the full\_leg\_2 viewpoint of the full-body dataset shown in \autoref{fig:heatmap}. In this case, participants in the user study slightly favored the hard clipping results. We attribute this to two key factors: (1) To ensure side-by-side comparison in our web-based user study interface, we downsampled the rendering resolution from 1024×1024 to 512×512. This resolution reduction likely caused the fine-grained differences, especially in boundary smoothness, to become less perceptible to users. (2) The Gaussians in the full-leg dataset are inherently small and compact. In such cases, both hard clipping and RaRa strategies yield similar visual results due to limited clipping overlap per primitive.

\textbf{Applicable scenarios}: RaRa is particularly suited for multi-layer Gaussian fields, where internal structures—not just surfaces—are of interest. While original Gaussian Splatting focuses on surface appearance, clipping enables volumetric exploration. As Gaussian-based representations evolve toward anatomy visualization or semantic layer decomposition, the need for precise and artifact-free clipping will grow, making RaRa increasingly relevant.

\textbf{Limitations:} Despite its generality and compatibility, our method currently supports only planar clipping surfaces. However, our method can be extended to support other clipping primitives, provided that efficient ray–primitive intersection tests are available. Moreover, our method operates only at the rendering level—the clipped result exists only during visualization and is not baked into the underlying data representation. Future work could explore methods to re-fit or re-optimize Gaussians post-clipping to persistently store editable or extractable subregions.
\section{Conclusion}
We presented RaRa Clipper, a novel clipping strategy for Gaussian Splatting that seamlessly integrates rasterization-based visibility classification with selective ray tracing. By applying ray–ellipsoid intersection only to Gaussians near the clipping boundary, and computing visibility-aware opacity using a decaying function, our method achieves high-fidelity, smooth, and temporally stable clipping without sacrificing real-time rendering performance. Extensive experiments and user studies confirm that our method outperforms the hard clipping strategy in both objective accuracy and perceptual quality. RaRa Clipper is plugin-compatible with existing GS renderers and provides an essential step toward precise, editable, and interactive Gaussian volume exploration.

\printbibliography

\clearpage

\end{document}